\documentclass[11pt]{iopart}
\usepackage{graphicx}
\def\bbh#1{binary black hole#1 (BBH#1)\gdef\bbh{BBH}}

\begin{document}

\title[Spinning Binary Black Holes]{Numerical Relativity meets Data Analysis: Spinning Binary Black Hole Case}
\author{Deirdre Shoemaker, Birjoo Vaishnav, Ian Hinder and Frank Herrmann}
\address{Center for Gravitational Wave Physics, Penn State,
University Park, PA 16802}

\begin{abstract}
  We present a study of the gravitational waveforms from a series of
  spinning, equal-mass black hole binaries focusing on the harmonic
  content of the waves and the contribution of the individual
  harmonics to the signal-to-noise ratio.  The gravitational waves
  were produced from two series of evolutions with black holes of
  initial spins equal in magnitude and anti-aligned with each other.
  In one series the magnitude of the spin is varied; while in the
  second, the initial angle between the black-hole spins and the
  orbital angular momentum varies.  We also conduct a preliminary
  investigation into using these waveforms as templates for detecting
  spinning binary black holes.  Since these runs are relativity short,
  containing about two to three orbits, merger and ringdown, we limit
  our study to systems of total mass $\ge 50 M_\odot$.  This
  choice ensures that our waveforms are present in the ground-based
  detector band without needing addition gravitational wave cycles.
  We find that while the mode contribution to the signal-to-noise
  ratio varies with the initial angle, the total mass of the system
  caused greater variations in the match.
\end{abstract}

\section{Introduction}
Gravitational waves produced during the coalescence of compact objects
such as black holes, are one of the most promising sources for
detection by interferometric detectors such as LIGO, VIRGO, GEO and
TAMA \cite{LIGO2006}.  For these ground based-detectors, where any
gravitational-wave signals may be deeply buried in detector noise, the
matched filtering technique \cite{Wainstein} is the detection strategy
of choice.  Matched filtering is optimal when accurate representations
of the expected signal are used.  For low mass compact object binaries
that means templates built from well-known analytic methods such as
the post-Newtonian (PN) approximation \cite{lrr-2006-4}.  When the
total mass is larger than approximately $50M_\odot$ \cite{FHI}, the
merger of the \bbh{} system will not only be present in the
sensitivity band of ground-based detectors but also generate the
strongest signal.  Quantifying the last orbits and merger of a \bbh{}
system has long been the purview of numerical relativity.

A new direction in the efforts to generate accurate templates is the
inclusion of waveforms produced by numerical relativity.  The
waveforms from numerical relativity may be used in data analysis
methods involving searches for inspiraling sources in several ways.
For instance, they can be used in creating hybrid templates for
detection, validating and extending the mass range of the current
search strategy and may be implemented as templates themselves
especially for the higher mass range \cite{buonanno-2006, Brady-2006,
  pan-2007,ajith-2007,vaishnav-2007,buonanno-2007,ajith-2007b}.

In a previous paper \cite{vaishnav-2007} (paper I), we studied
waveforms from a series of numerical evolutions of equal-mass,
spinning \bbh{} mergers, which we label the A-series.  In the
A-series, the black holes had spins that were equal in magnitude with
one spin aligned and the other anti-aligned with the orbital angular
momentum, $J_{orb}$.  We found that using only the dominant harmonic
of the radiation, a common practice in numerical relativity and data
analysis, caused a complete degeneracy of the this spin space with
respect to detection.  In other words, a template of a non-spinning
\bbh{} waveform faithfully matched the signal of any anti-aligned
spinning waveform when only the dominant modes were used.  However,
retaining non-dominant harmonics of non-negligible amplitude broke the
degeneracy to some degree, most notably for spins $J/M^2 \ge 0.6$.  In
this paper, we further our previous study in two ways. First we
include a new series of \bbh{} mergers, the B-series, in which the
initial black-hole spins are still equal in magnitude and anti-aligned
with each other but the initial angle the black-hole spins make with
the orbital angular momentum, $\vartheta$, is allowed to vary.
Second, for both the A and B series of data we conduct a new study of
the contributions of the individual modes to the signal-to-noise
ratio.  In addition, for the B-series we also calculate the dependence
of the minimax matches between the dominant mode and the full waveform
on the initial spin orientation.

\section{The Waveforms}
The binary black hole coalescence problem can be divided into three
phases called the inspiral, merger and ringdown.
Due to the early accessibility of PN and other analytic
based approaches, most of the work in setting up detection schemes has
been done with these analytic approaches.  These methods are well
suited for the inspiral phase of the coalescence, only breaking down
at some yet-to-be-determined point within a few orbits of the merger.
Fortunately, the signal resulting from the inspiral is in the
ground-based frequency band for systems of total mass less than
approximately $50 M_\odot$.  As the mass increases, the inspiral
lowers in frequency, and the detectable signal contains more merger
and ringdown.  Now that numerical relativity is producing the
waveforms for the final orbits and the merger phase of the coalescence
\cite{Pretorius:2005gq,CampanelliPRL2006,Baker05a} we can investigate
the inclusion of the merger regime in detection strategies.

We study two sets of waveforms both the result of evolutions conducted
by the PSU numerical relativity group.  The A-series was published in
\cite{Herrmann07} with initial black-hole spins covering the set
$a={0.0,0.2,0.4,0.6,0.8}$.  The B-series was published in
\cite{Herrmann07b} and generalizes the A-series with variation of the
initial angle that one of the anti-aligned spins makes with the axis,
$\vartheta$, at a fixed magnitude of spins, $a=0.6$.  When
$\vartheta=0$, we recover the a=0.6 waveform of A-series.  When
$\vartheta=\pi/2$, the spin-directions lie in the plane of the orbit
and are in the``superkick'' \cite{gonzalez-2007} configuration in
which the maximum gravitational recoil from the \bbh{} mergers has
been found.

Since these waveforms were originally produced to study the
gravitational recoil imparted to the final black hole after an
asymmetric collision, only two to three orbits were evolved (the
merger phase dominates the recoil).  The number of orbits is set by
the initial orbital frequency for a given total mass.  In order to
place the numerical waveforms firmly in the frequency band of the
detector, we use the initial LIGO noise curve \cite{LIGO:SRD} and we
only investigate masses larger than $50 M_\odot$ when calculating
matches between waveforms.  The total mass sets the frequency at which
the signal enters the band.  For example, the cutoff frequency for a
binary system of $50 M_\odot$ is $0.02/M$ or approximately $80$Hz and
about $40$Hz for $100 M_\odot$.

The waveforms were extracted from the numerical evolution of the
spacetime in terms of the Newman-Penrose scalar, $\Psi_4(t,x,y,z)$,
which is expanded into angular modes via ${}_{-2}Y_{\ell
  m}(\theta,\phi)$, the spin-weighted $s=-2$ spherical harmonics, by
extraction on a sphere. The dominant mode for the quasi-circular
orbits is the quadrupole mode ($\ell=m=2$).  The angles
$\theta$ and $\phi$ correspond to the inclination and azimuthal angles
between the source and detector in the source frame. When $\theta=0$,
the observer is directly above the orbital plane of the binary and
sees primarily the $\ell=m=2$ mode.  As $\theta$ increases, the
waveforms are a mixture of modes.  We truncate the infinite series of
modes at $\ell \leq 4$ because we do not extract all the modes from
the simulations and modes of $\ell>4$ were zero within our numerical
error.  As more complicated configurations are evolved, more modes
will need to be accurately extracted from the codes.

\section{Faithfulness}
The multipolar analysis of \bbh{} waveforms produced by numerical
relativity has been pursued for both unequal mass and spinning \bbh{}
configurations \cite{schnittman-2007,BertiJena,BertiAmaldi}.  In paper
I, we found that using only the dominant mode in comparing waveforms,
tantamount to choosing an inclination angle with the detector in the
source frame of $\theta=0$, resulted in a degeneracy of the A-series
parameter space.  In this paper, we focus our attention on how
different initial configurations, in this case $a$ and $\vartheta$,
result in different mode contributions to the signal-to-noise ratio.
To build intuition about what parameters might be important to the
template space of black-hole mergers, we further calculate the overlap between
pairs of our waveforms.  We perform a preliminary analysis on how
faithful $\ell=m=2$ waveforms of various parameters would be in
matching with waveforms at random inclination angle.  We keep the
total mass for each template fixed and vary the inclination angle of
the detector, $\theta$, the spin $a$, and initial angle $\vartheta$ when appropriate.

The minimax match is given by \cite{Owen-96,damour-1998-57}
\begin{equation}
M \equiv  \max_{t_0}  \min_{\Phi} \frac{ \langle h_1 |h_2
\rangle} {\sqrt{\langle h_1 | h_1 \rangle \langle h_2 | h_2 \rangle
}}\,,
\label{eqn:minimax}
\end{equation}
where
\begin{equation}
\label{eq:inner}
\langle h_1 | h_2 \rangle = 4 \, \mbox{Re}\int_{f_{\mathrm{min}}}^{f_{\mathrm{max}}} \frac{\tilde
  h_1(f) \tilde h^*_2(f)}{S_{h}(f)} \, df.
\end{equation}
The Fourier transform of the strain, $h(t)$, is given by $\tilde h_+
(f) = \mathcal F(\mathrm{Re}(\Psi_4))(f)/(-4 \pi^2 f^2) \,$ where
$\Psi_4(t) = \frac{d^2}{d t^2}(h_+(t)-ih_\times(t)) \,,  $ and
$\mathcal F$ is a Fast Fourier Transform.  The signal-to-noise ratio,
$\rho$, is given by
\begin{equation}
\rho= \left[4 \int_{f_\mathrm{min}}^{f_\mathrm{max}} \frac{|\tilde h(f)|^2}{S_h(f)} \right]^{1/2}\,.
\end{equation}
The variable $S_h(f)$ denotes the noise spectrum for which we use the
initial LIGO noise curve.  The domain
$[f_{\mathrm{min}},f_{\mathrm{max}}]$ is determined by the detector
bandwidth and the masses of our signal.  The masses are set such that
the overlaps will not change significantly if we were to add the
inspiral portion of the signal because the A and B series of waveforms
have orbital frequencies that increase almost monotonically. Owing to
this, the gravitational wave frequency also increases monotonically
with time implying that extending the signal back in time will not
change the spectrum in the merger band. When precession is significant
this will no longer be true and the inspiral will likely contribute to
the signal at higher frequencies.

\begin{figure}[h]
\centerline{
\includegraphics[height=5cm]{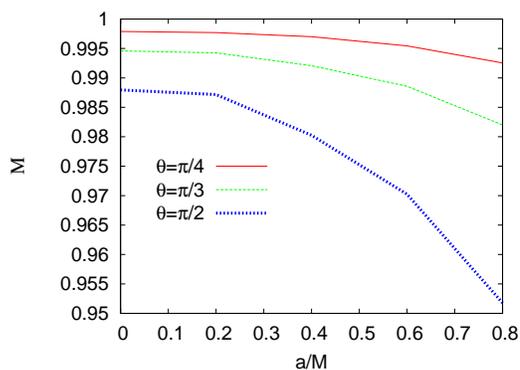}
}
\caption{The minimax match versus $a$ for three values of $\theta$,
  ${\pi/4, \pi/3, \pi/2}$ for the case of total mass $100 M_\odot$.
  The value of $\theta=0$ is not included in the plot, since this is
  the waveform of comparison and the match is one by definition.}
\label{fig:degenerateA-series}
\end{figure}

\noindent {\bf A-series:} In Fig.~\ref{fig:degenerateA-series} we plot
the match versus spin at different inclination angles, $\theta$, for a
given total mass of $100 M_\odot$.  This plot first appeared in paper
I and is included here for reference. The figure shows that as the
spin increases, the match between a waveform of $\theta=0$ and one of
non-zero $\theta$ decreases.  This indicates that the non-dominant
modes are important both for distinguishing between different spinning
waveforms and in making a detection.  This is most notable for the
$a=0.8$ case.

\begin{figure}[h]
\vbox{
\hbox{
\includegraphics[height=5cm]{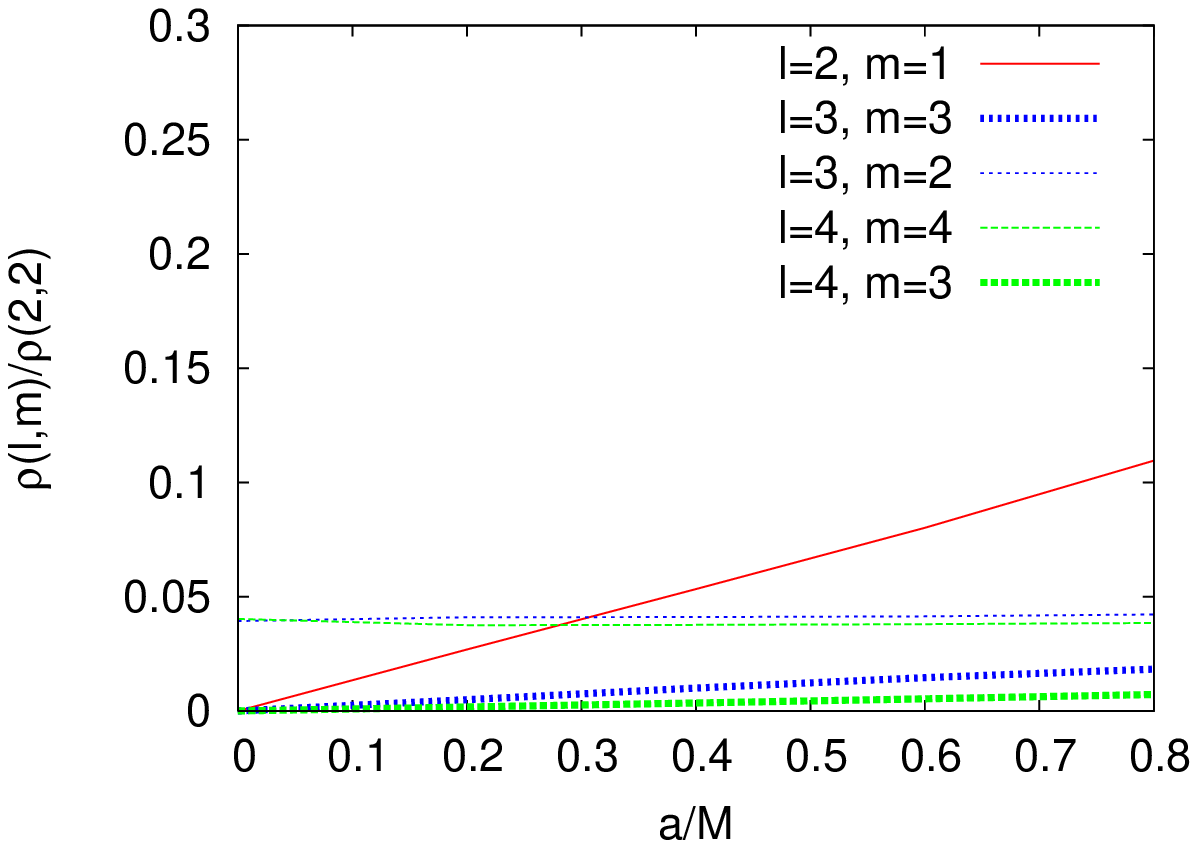}
\includegraphics[height=5cm]{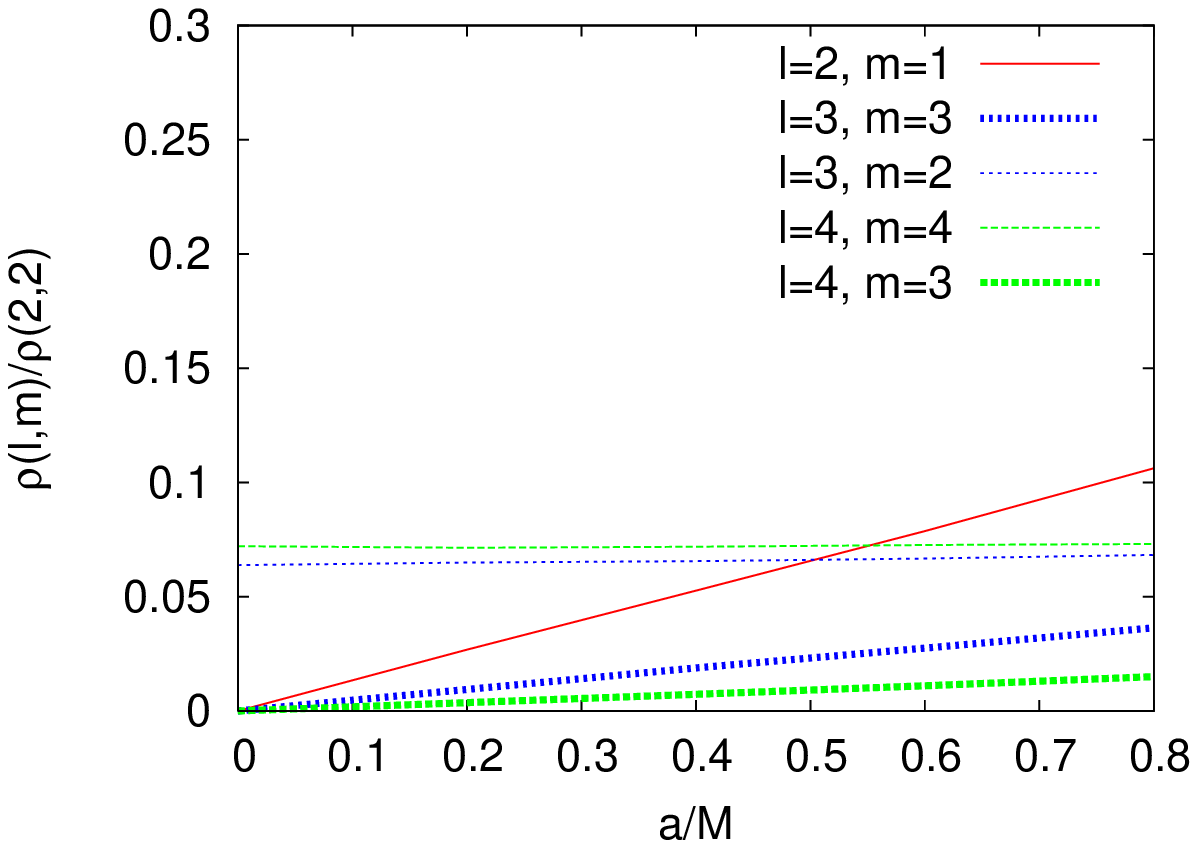}
}
\hbox{
\includegraphics[height=5cm]{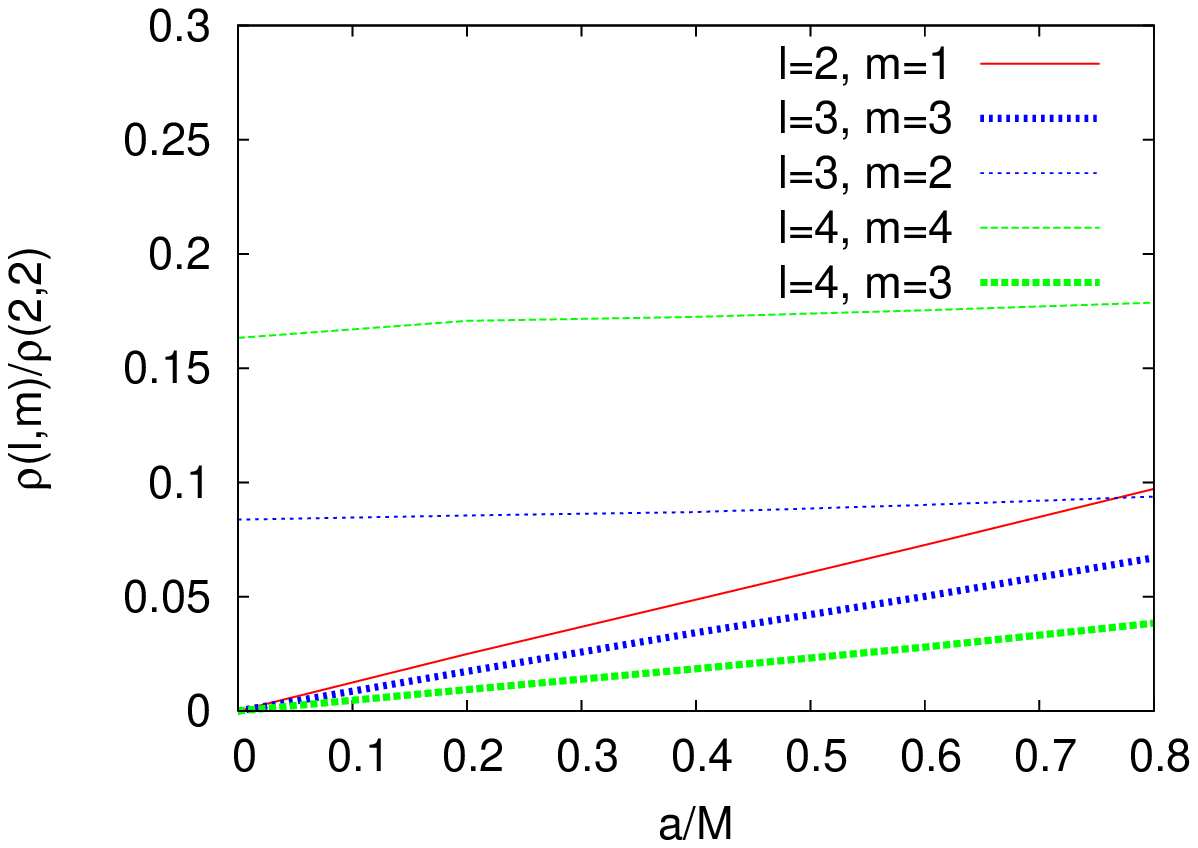}
\includegraphics[height=5cm]{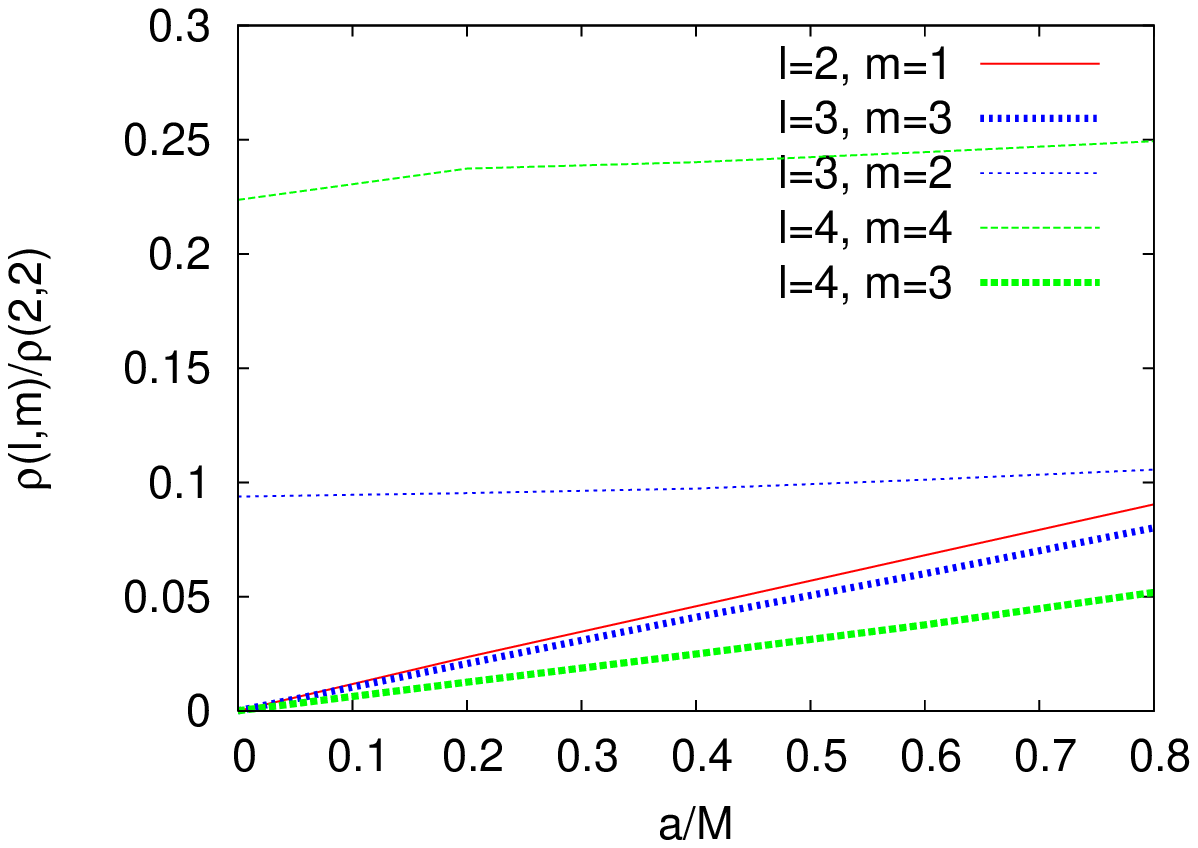}
}
}
\caption{A series of plots are shown with the ratio of $\rho$ per mode
  to the $\rho$ of the $\ell=m=2$ mode versus the initial spin of the
  black holes.  This is done for series-A.  Each plot refers to the
  calculation for a different total mass of the binary.  Starting from
  the upper left and moving right and then down, we have
  $50 M_\odot$, $100 M_\odot$, $200 M_\odot$, and
  $300 M_\odot$ on the lower right.}
\label{fig:snrvsspin}
\end{figure}

In order to have a qualitative understanding of why the match behaves
as in Fig.~\ref{fig:degenerateA-series}, we study the $\rho$ per mode
for the A-series.  This is only a qualitative estimate because the
relative fraction of modes present in the signal will depend on the
relative spin-weighted spherical harmonics values at the particular
angle. In practice the error induced by ignoring the mixed terms that
are important in constructing $\rho(\theta,\phi)$ from the modes is
less than 20\%, as the relative overlaps of the significant modes from
both the A and B-series of data are of this order.  Since the $\rho$
of the $\ell=m=2$ mode is much larger than the $\rho$ of the
other modes, we plot the ratio, $\rho(\ell,m)/\rho(2,2)$ in
Fig.~\ref{fig:snrvsspin}.  The upper left plot corresponds to a
system of mass $50 M_\odot$, the upper right to $100 M_\odot$, lower
left to $200 M_\odot$ and lower right to $300 M_\odot$.
Across all the masses sampled, the ratio of the $\rho$ for each mode
grows with increasing spin. This is especially true for the $m=1$ and
$m=3$ modes which are suppressed at low $a$.  The $m=2$ and $m=4$
increase slightly with $a$.  While the $\ell=2$, $m=1$ mode is the
next mode dominant mode after the $\ell=m=2$ mode for the high-spin
regime at low masses, the $\ell=m=4$ mode is secondary for the entire
$a$ range at higher masses.  For low spins, the waveform is entirely
dominated by the $m=2$ modes. The linear-like growth of the odd-$m$
modes with spin magnitude is expected from PN expressions like Eq(1)-(4)
in \cite{BertiAmaldi}.

\begin{figure}[h]
\centerline{
\includegraphics[height=5cm]{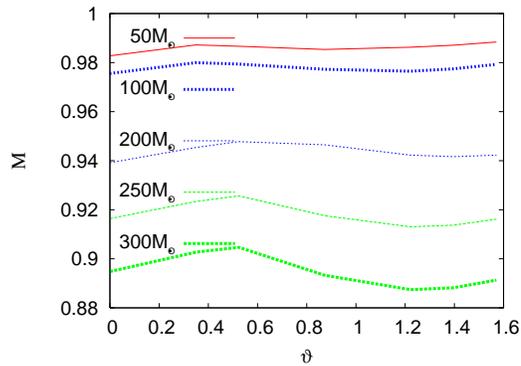}
}
\caption{The minimax match versus initial angle $\vartheta$ for the B-series.
  Each curve represents a particular choice of total mass, $50
  M_{\odot}$ at the top with each successively lower line a higher
  mass.  To compute the match, we used one waveform with $\theta=0$
  and the other at $\theta=\pi/2.4$ radians for a given $a$ and
  $\vartheta$.}
\label{matchvsentrance}
\end{figure}

\noindent {\bf B-series:} The minimax match versus the initial angle
for the B-series is presented in Fig.~\ref{matchvsentrance}.  Each
line represents a choice of total mass, with $50 M_\odot$ the top most
line and $300 M_\odot$ the bottommost.  The match was computed by
setting one waveform to $\theta=0$ and the other to $\theta =
\pi/2.4$.
We find that the variation of the match across initial angle for the
given spin of $a=0.6$ does not change more than about $2\%$. The
variation amongst different total masses is more dramatic, dropping
down below a match of 0.9 for most in the angles at a mass of $300
M_\odot$.  At that large mass range, the ringdown is contributing
significantly to the signal, and differences in the modes, like the
$\ell=m=4$ mode, begin to make important contributions.  These \bbh{}
configurations settle down to a final black hole with a spin of
$a=0.62$.

\begin{figure}[h]
\vbox{
\hbox{
\includegraphics[height=5cm]{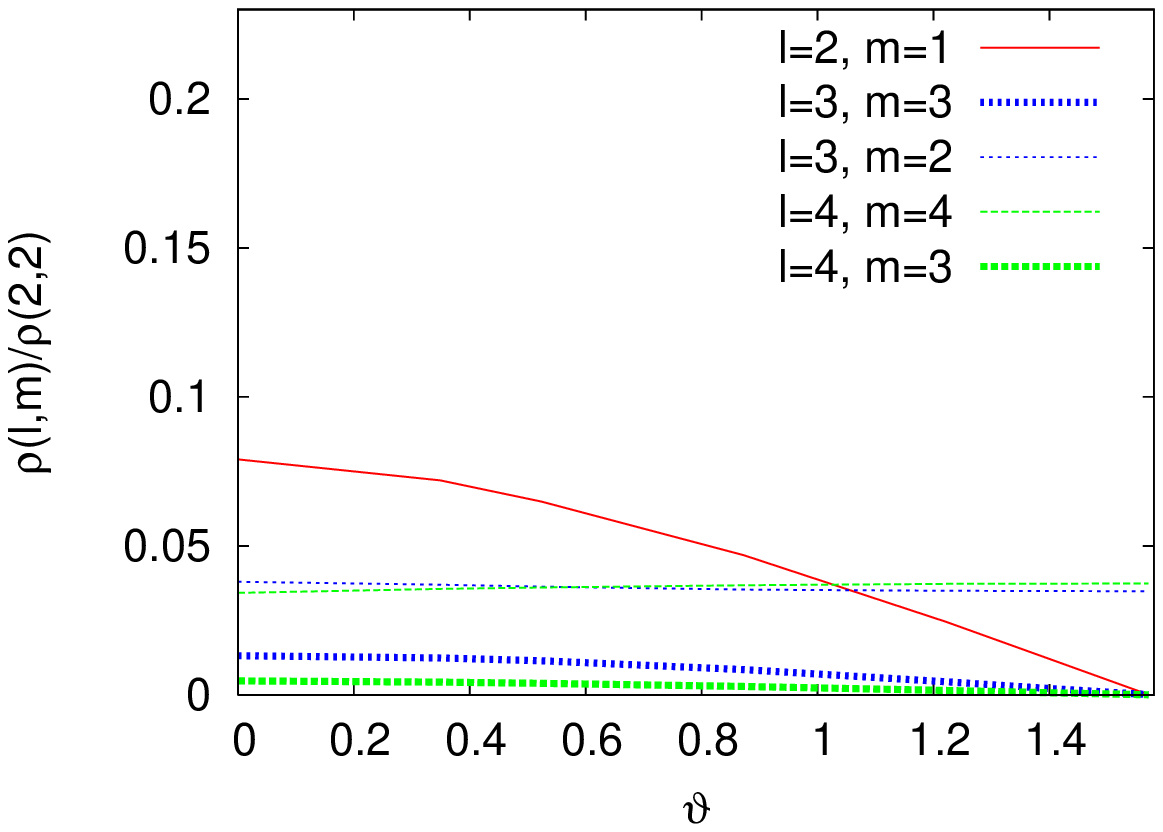}
\includegraphics[height=5cm]{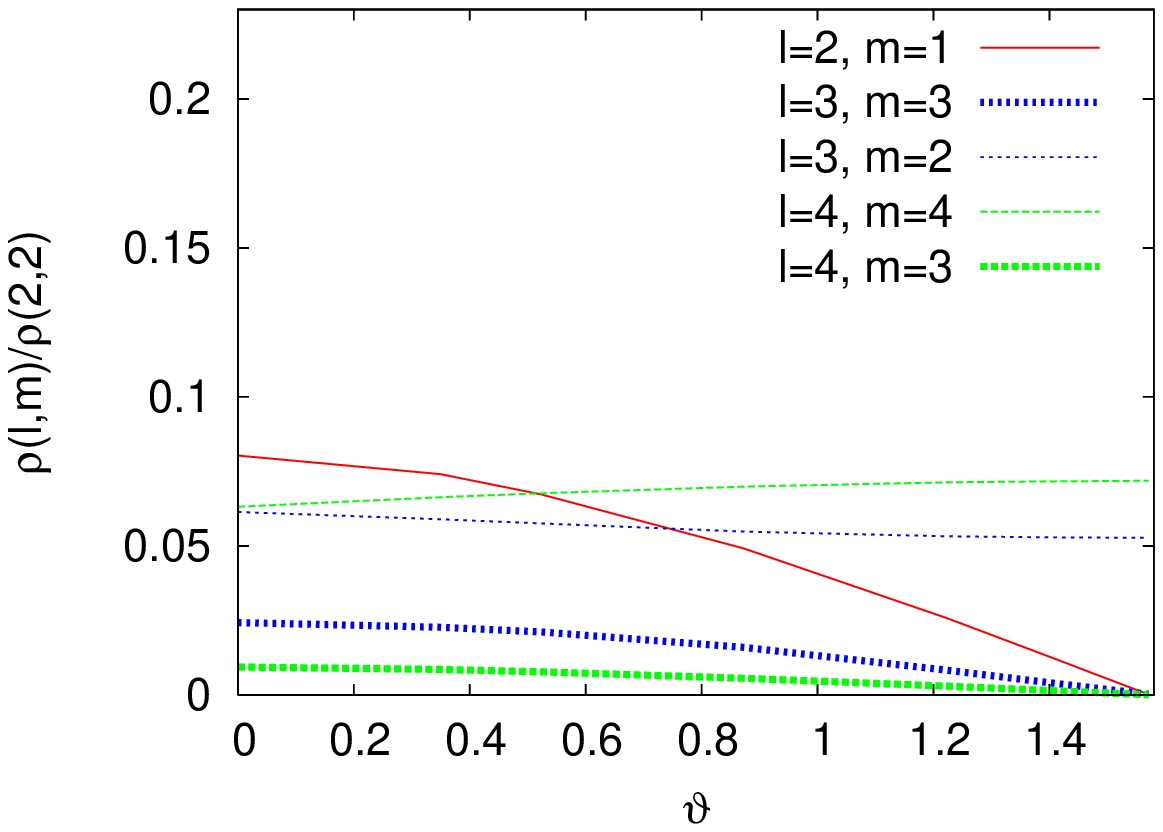}
}
\hbox{
\includegraphics[height=5cm]{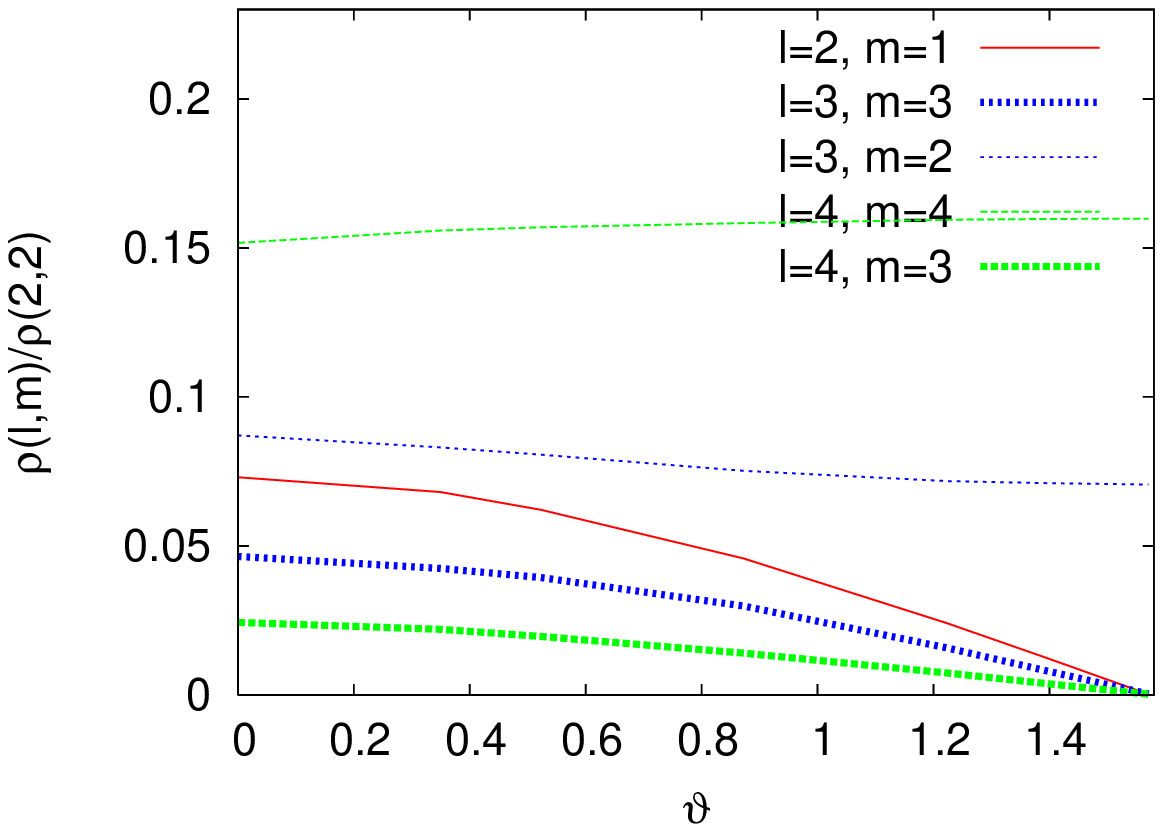}
\includegraphics[height=5cm]{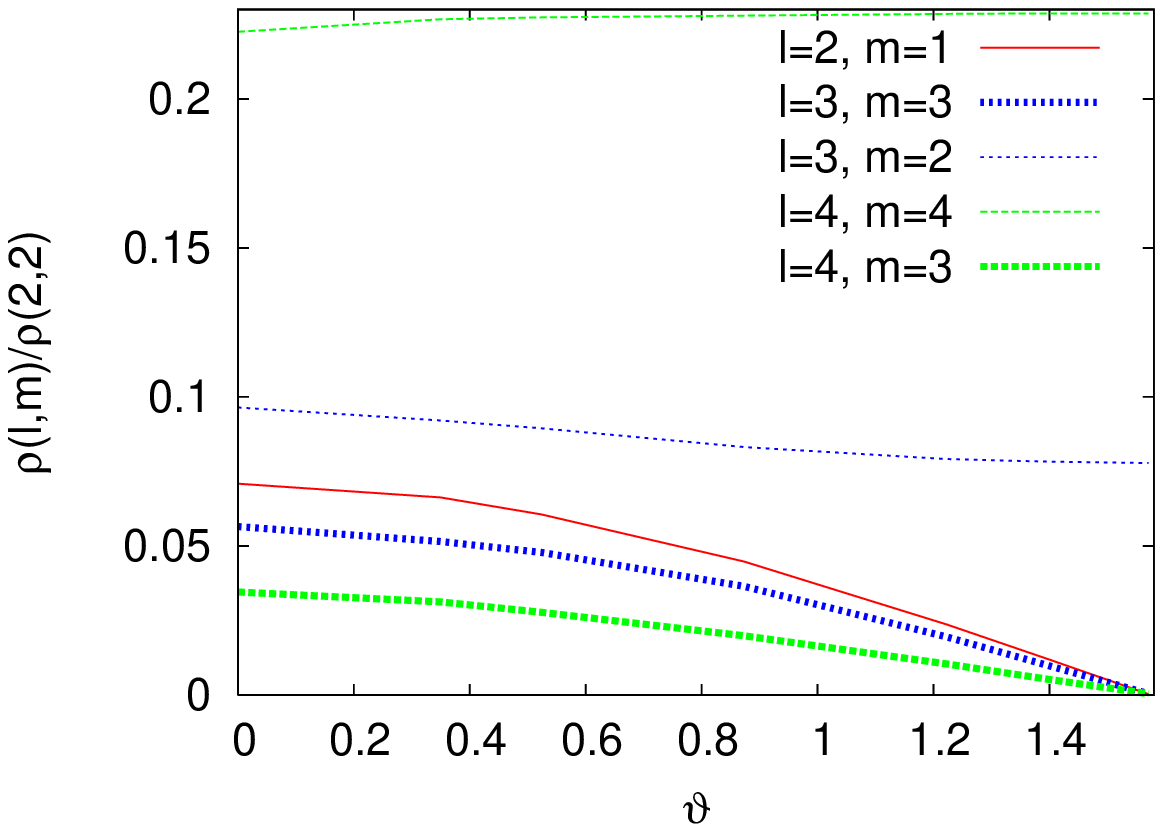}
}
}
\caption{A set of plots is shown with the ratio of $\rho$ per mode to
  the $\rho$ of the $\ell=m=2$ mode versus the initial angle for
  series-B.  Each plot is the ratio computed for a different total
  mass of the binary.  Starting from the upper left and moving right
  and then lower left and right, we have $50 M_\odot$, $100 M_\odot$,
  $200 M_\odot$, and $300 M_\odot$ on the lower right.}
\label{fig:snrvsangle}
\end{figure}

In Fig.~\ref{fig:snrvsangle}, we once again investigate a qualitative
interpretation of the match through the $\rho$
as plotted versus the initial angle, $\vartheta$ for each mode.
The upper left plot corresponds to a system of mass $50 M_\odot$, the
upper right to $100 M_\odot$, lower left to $200 M_\odot$ and lower
right to $300 M_\odot$.
Across the mass scales sampled, as $\vartheta$ increases, the signal in
odd-$m$ modes decrease. In the non-precessing case, the strength of
the odd-modes is expected to vary with the z-component of the spin
\cite{BertiAmaldi}.  Since in this series of runs, the spins precess
about the z-axis and $\vartheta$ remains nearly constant, the relative
strength of the modes show similar trends as the non-precessing case.
At higher masses, where the ringdown dominates the signal, the
$\ell=m=4$ mode contributes a large portion of the $\rho$ of the total
signal, over 20\% for a \bbh{} of $300 M_\odot$.  This is in part due
to the enhancement of the ringdown signal which occurs when the
frequency of the higher modes lies around the detector's sweet spot.
It is interesting to note that in the "superkick" the spread of the
modes is reduced in $\rho$ compared with the parallel configuration at
$\vartheta=0$.  The decrease of the odd $m$ modes is expected from the
PN expressions.  For example, the $\ell=2$, $m=1$ will be suppressed
when the spins lie in the orbital plane for equal-mass black holes as
discussed in \cite{BertiAmaldi}.

These waveforms are the solution to the \bbh{} coalescence as
expressed by general relativity with errors arising from several
sources, see \cite{Boyle:2007ft}.  In paper I, we analyzed the effects
of resolution on the matched filtering technique and found that for
the resolutions used to compute the waveforms studied in the A-series,
the largest error would be $\pm 0.02$ in the match, although that is
only for the $a=0.8$ case, and is typically smaller.  The waveforms in
the B-series have comparable errors, i.e. the resolutions, wave
extraction and other numerical techniques were the same in computing
both series of waveforms as discussed in
\cite{vaishnav-2007,Herrmann07,Herrmann07b}.  For reference, the
typical resolution on the finest grids were $h=M/35.2$ where $M$ is
the total mass.

\section{Discussion and Conclusion}
In this paper, we investigated the contribution of individual modes to
$\rho$ from the last orbits, merger and ringdown of an equal-mass,
spinning binary black coalescence.  In the A-series, the
spins were kept parallel/anti-parallel to the direction of the orbital
angular momentum, but the magnitude of the spins varied.  In the
B-series, the magnitude was kept fixed to $a=0.6$, but the initial
angle the spins make with the orbital angular momentum varied.

In paper I, we investigated the match between a waveform from the
A-series containing only the $\ell=m=2$ mode and a waveform of a sum
of modes.  There we found strong dependence on the match with spin,
with the $\ell=m=2$ waveform failing to match to spinning waveforms
especially for spins equal to and greater than $a=0.6$.  We did a
similar study here for the B-series, comparing two waveforms of
$a=0.6$ at various $\vartheta$.  We found, despite the variation of
the $\rho$ versus $\vartheta$, the match had a much greater dependence
on mass than the initial angle.  The inclusion of modes was much more
important to templates of higher mass, where the merger and ringdown
dominate the signal, than at lower masses.  This importance will be
more evident in matches using unequal-mass and spinning waveforms with
larger spin as well as waveforms with more cycles.

To qualitatively understand the matches, we conducted a multipolar
analysis of the modes in each waveform and calculated the ratio of the
$\rho$ of each mode versus $\ell=m=2$.  For the A-series, the $\rho$
per mode increased as the magnitude of the spins increased at every
total mass as seen in Fig.~\ref{fig:snrvsspin}.  The odd-$m$ modes
increased from almost no contribution at low spins to a $10\%$
contribution at larger spins.  At a given spin, the $\ell=2, m=1$ mode
dominated the $\rho$ at low mass, but the $\ell=m=4$ mode's ratio to
$\ell=m=2$ grew with increasing mass.  For the B-series,
Fig.~\ref{fig:snrvsangle}, we found that the diversity of contributing
modes decreases with increasing angle, except for the $\ell=m=4$ and
$\ell=3$, $m=2$ modes which remain relatively constant across
$\vartheta$ for a given mass.  As in the variation with $a$, at low
total binary mass, the secondary signal is the $\ell=2$, $m=1$ mode,
but at higher masses the $\ell=m=4$ and the $\ell=m=2$ modes are
stronger.  As anticipated, the $\ell=2$, $m=1$ mode decreases to zero
as the initial angle moves to lie parallel to the orbital plane.  For
both the series of runs, the variation of the signal in different
modes is consistent with the expectation from PN \cite{BertiAmaldi}.

\section{Acknowledgments}
We thank The Center for Gravitational Wave Physics is supported by the
NSF under cooperative agreement PHY-0114375. Support for this work was
also provided by NSF grants PHY-0653443 and PHY-0653303.

\bibliographystyle{unsrt}
\bibliography{refs}

\end{document}